\documentclass[prl,nopacs,twocolumn,preprintnumbers,notitlepage,amsmath,amssymb,superscriptaddress]{revtex4-1}

\usepackage{color}
\usepackage{pgffor}
\usepackage[T1]{fontenc}
\usepackage[pdftex]{graphicx}
\usepackage[]{pdfpages}

\makeatletter
\AtBeginDocument{\let\LS@rot\@undefined}
\makeatother

\definecolor{jerem}{rgb}{1, 0, 0}
\definecolor{jerem2}{rgb}{0, 0, 1}
\definecolor{olivier}{rgb}{0.125, 0.26, 0.07}

\usepackage{makecell}

\usepackage{enumerate}
\usepackage{mathtools}
\usepackage{stmaryrd}

\usepackage{enumitem}

\usepackage{epstopdf}

\usepackage{textcomp}
\usepackage{gensymb}
\def \equi#1{\mathrel{\mathop{\kern 0pt\sim}\limits_{#1}}}

\begin{document}
\title{Splitting Probabilities of Jump Processes}
%\author{Jeremie Klinger et Muriel Livernet}
\author{J. Klinger}
\affiliation{Laboratoire de Physique Th\'eorique de la Mati\`ere Condens\'ee, CNRS/Sorbonne Université, 
 4 Place Jussieu, 75005 Paris, France}
\affiliation{Laboratoire Jean Perrin, CNRS/Sorbonne Université, 
 4 Place Jussieu, 75005 Paris, France}
\author{R. Voituriez}
\affiliation{Laboratoire de Physique Th\'eorique de la Mati\`ere Condens\'ee, CNRS/Sorbonne Université, 
 4 Place Jussieu, 75005 Paris, France}
\affiliation{Laboratoire Jean Perrin, CNRS/Sorbonne Université, 
 4 Place Jussieu, 75005 Paris, France}
 \author{O. B\'enichou}
\affiliation{Laboratoire de Physique Th\'eorique de la Mati\`ere Condens\'ee, CNRS/Sorbonne Université, 
 4 Place Jussieu, 75005 Paris, France}
 
%\date{\today}

\begin{abstract}

We   derive a  universal, exact asymptotic form of the splitting probability for  symmetric continuous jump processes, which quantifies the  probability $ \pi_{0,\underline{x}}(x_0)$ that the process crosses $x$ before 0 starting from a given position $x_0\in[0,x]$ in the regime $x_0\ll x$. This analysis   provides in particular a fully explicit determination of the transmission probability ($x_0=0$), in striking contrast with the trivial  prediction $ \pi_{0,\underline{x}}(0)=0$ obtained by taking  the continuous limit of the process, which  reveals  the importance of the microscopic properties of the dynamics. These results are illustrated with paradigmatic models of jump processes with applications to light scattering in heterogeneous media in realistic 3$d$ slab geometries. In this context, our explicit predictions of the transmission probability, which can be directly measured experimentally,  provide a  quantitative characterization of the effective random process describing light scattering in the medium.

\end{abstract}
\date{\today}

\maketitle

The splitting probability quantifies the  likelihood of a  specific outcome out of several alternative possibilities for a random process \cite{Redner:2001a,Kampen:1992,Hughes:1995,Gardiner:2004}. While these quantities can be defined for general $d$-dimensional stochastic processes and any number of possible outcomes \cite{Condamin:2008,al:2011}, most examples of applications concern 1-dimensional processes with two outcomes; one then defines  $\pi_{0,\underline{x}}(x_0)$  as the probability that the process crosses $x$ before 0 starting from $x_0$. A celebrated example is given by the Gambler's ruin problem \cite{Redner:2001a}, schematically quantified by the splitting probability that a 1-dimensional random walker (figuring the gambler's fortune) reaches 0 (complete ruin) before a fixed given threshold; other examples  are given by the fixation probability of a mutant in the context of population dynamics \cite{moran}, or the melting probability of a heteropolymer \cite{Oshanin_2009}, which can be re-expressed in terms of splitting probabilities. A key example, to which we will refer through this paper, is given by the transmission probability of particles (eg photons or neutrons) through a slab of a scattering medium, which has important applications in various fields \cite{Rotter:2017uy,Burioni:2010tr,PhysRevE.89.022135,PhysRevE.90.052114,PhysRevE.103.L010101}; in this case the transmission probability is nothing but the splitting probability for the particle to reach the exit side rather  than being back-scattered. 
\begin{figure}[h!]
    \centering
    \includegraphics[scale=0.5]{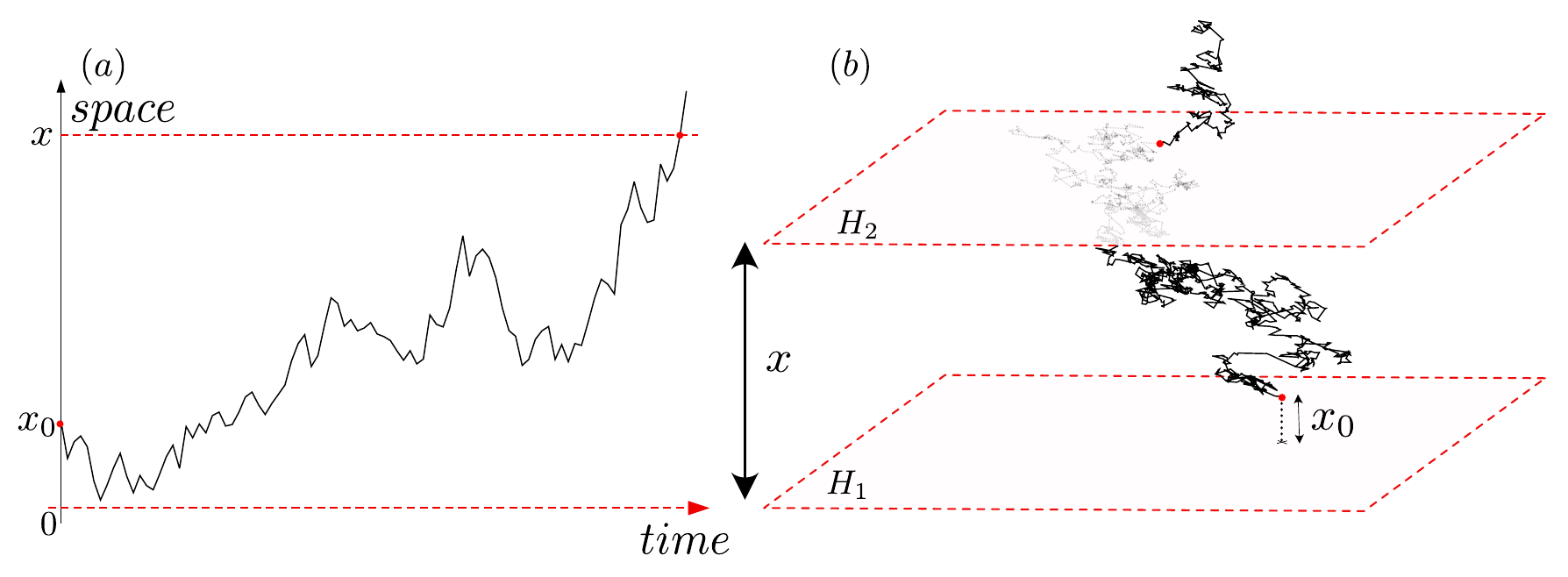}
    \caption{(a) the one dimensional jump process evolves in the bounded interval $[0,x]$. One is interested in the probability $\pi_{0,\underline{x}}(x_0)$ of crossing $x$ before 0 starting from $x_0$ ,as shown in the diagram. (b) The jump process is now evolving in a 3 dimensionnal space and is bound to stay between two hyperplanes $H_1$ and $H_2$ distant of $x$. We now want to evaluate the probability of crossing $H_2$ before $H_1$, starting from $x_0$.}
    \label{fig : schematics}
\end{figure}

There is to date no explicit determination of the splitting probability for general jump processes \cite{Kampen:1992,Levernier:2021aa,MAJUMDAR20104299}. Jump processes are defined as follows for $d=1$: at each discrete time step $n$, the walker performs a jump of extension $l\in {\mathbb R}$ drawn according to a distribution $f(l)$ whose Fourier Transform will be denoted $\tilde{f}(k) = \int_{-\infty}^\infty e^{ikl}f(l)\text{d}l$. For jump processes, the splitting probability is known to satisfy the following integral backward equation \cite{Kampen:1992}:
\begin{equation}\label{int}
\pi_{0,\underline{x}}(x_0)=\int_{x-x_0}^\infty dx' f(x') + \int_{-x_0}^{x-x_0} dx' \pi_{0,\underline{x}}(x_0+x')f(x')
\end{equation}
which results from a partition over the first jump.  Even if this equation is linear,  there is to date no available  solution with the exception of the exponential distribution $f(l)= e^{-|l|/\gamma}/(2\gamma)$;   the main difficulty lies in the finite integration range, which prevents the use of classical integral transforms \cite{Majumdar:2006aa}.

 An important simplification of the problem  is achieved by taking a continuous limit. For symmetric jump processes, considered in what follows, the small $k$ expansion of $\tilde{f}(k)$ reads
 \begin{equation}\label{fk}
 \tilde{f}(k) \underset{k\rightarrow 0}{=} 1 -(a_\mu |k|)^\mu + o(k^\mu)
 \end{equation}
where $a_\mu$ defines the microscopic characteristic length scale of the process. Two limit behaviors emerge \cite{Bouchaud:1990b,R.Metzler:2000}. For $\mu= 2$ the variance of the jump distribution is finite and the process is known to converge at large times to Brownian motion; for $0<\mu< 2$, the process converges instead to an $\alpha$-stable Levy process of parameter $\mu$. Hence, there are three independent length scales in the problem: $x_0,x,a_\mu$, which can lead to two distinct asymptotic regimes. Taking $a_\mu\ll x_0<x$ defines   the continuous limit of the problem \eqref{int}, whose solution can  be obtained and reads \cite{Widom1961,Blumenthal1961,Majumdar:2010}: 
\begin{equation}\label{Zoia}
    \pi_{0,\underline{x}}(x_0) = \frac{\Gamma(\mu)}{\Gamma^2(\frac{\mu}{2})}\int_0^{\frac{x_0}{x}}\left[u(1-u)\right]^{\mu/2-1} \text{d}u.
\end{equation}
The regime  $x_0 \ll x $ is of particular interest and has received a marked attention \cite{Majumdar:2010}.  One obtains from \eqref{Zoia} that this regime is given by \footnote{Remarkably, this  scaling behavior with $x_0/x$ has been generalized to non Markovian  scale invariant processes, at the cost of determining the persistence exponent $\theta$.} :
\begin{equation}\label{eq : small_x0}
    \pi_{0,\underline{x}}(x_0)\underset{a_\mu\ll x_0\ll x}{\sim}\frac{2\Gamma(\mu)}{\mu\Gamma^2(\frac{\mu}{2})}\left(\frac{x_0}{x}\right)^{\frac{\mu}{2}}.
\end{equation}
As explained above, a key application of splitting probabilities is the determination of the transmission probability of particles through a slab, that can be defined as $ \pi_{0,\underline{x}}(x_0=0)$.  The blunt use of the continuous limit  \eqref{eq : small_x0} yields   $ \pi_{0,\underline{x}}(x_0=0)= 0$, in clear contradiction with the expected result 
for  a  jump process with finite microscopic length scale $a_\mu$, for which      $\pi_{0,\underline{x}}(x_0=0)>\int_{x}^\infty dx' f(x')>0$.
Finally, the determination of the transmission probability requires to consider the second, distinct regime $x_0\ll a_\mu$ and thus  to go beyond the continuous limit \eqref{eq : small_x0}; this is the main purpose of this letter. 

Jump processes with finite microscopic length scale $a_\mu$ have proved to be relevant in various contexts \cite{Ziff:2009vh}. They provide emblematic models  of transport of photons or neutrons  in scattering media \cite{Rotter:2017uy}. More recently, they have  gained renewed interest in the context of self-propelled particles, be them artificial or living, such as active colloids, cells or larger scale animals \cite{Romanczuk:2012fk,Tejedor:2012ly,Levernier:2021aa,Meyer:2021tx,Mori2020a}. In what follows, we derive a  universal form for the splitting probability for  continuous jump processes of finite length scale $a_\mu$ in the regime $x_0\ll a_\mu \ll x$, which provides in particular an explicit determination of the transmission probability ($x_0=0$), and reveals  the importance of the microscopic properties of the process. These results are illustrated with paradigmatic models of jump processes with applications to light scattering in heterogeneous media.

{\it General results.} We first derive an asymptotic   expression of the splitting probability $ \pi_{0,\underline{x}}(x_0)$ for general $1d$ continuous symmetric jump processes of characteristic microscopic length scale $a_\mu$ as defined above  in the limit $x\to \infty$.   Denoting $F_{\underline{0},x}(n|x_0)$  the probability that the process starting from $x_0\in [0,x]$ crosses $0$ before $x>0$ for the first time after exactly $n$ steps, and making a partition  over the crossing time yields  : 
\begin{equation}\label{eq : time_partition}
   1- \pi_{0,\underline{x}}(x_0)\equiv\pi_{\underline{0},x}(x_0) = \sum_{n=1}^\infty F_{\underline{0},x}(n|x_0).
\end{equation}
This exact equation expresses the splitting probability in terms of two targets first-passage time distributions $F_{\underline{0},x}(n|x_0)$, for which no explicit solutions are available for general jump processes. Adapting the approach   introduced for scale invariant processes in $1d$  \cite{Majumdar:2010} and then extended to $d-$dimensional compact cases  \cite{Levernier:2018qf},   we next show that in the asymptotic limit $x\to \infty$, the splitting probability of jump processes  can in fact be re-expressed in terms of one target first-passage time distributions. We first note that  in \eqref{eq : time_partition} the right hand side involves trajectories that cross $0$ before $x$; most of these events thus occur within the typical number of steps  $n_{typ}$ needed to cross  $x$. In the regime $x\gg a_\mu,x_0$, we argue that $n_{typ}$ is simply the timescale to cover a distance $x$ \cite{Bouchaud:1990b} and thus satisfies $n_{typ} \sim \alpha  x^{\mu}$ where $\alpha$ is a process dependent constant (independent of $x_0$). We next remark that for time scales $n<n_{typ}$, the target at $x$ is  irrelevant so that $F_{\underline{0},x}(n|x_0)\simeq F_{\underline{0},\infty}(n|x_0)$, which leads to 
\begin{equation}\label{eq : time_partition_v2}
    \pi_{\underline{0},x}(x_0) \sim \sum_{n=1}^{n_{typ}} F_{\underline{0},\infty}(n|x_0)\equiv 1-q(x_0,n_{typ})
\end{equation}
where $q(x_0,n)=\sum_{k=n+1}^{\infty} F_{\underline{0},\infty}(k|x_0)$ is the survival probability, \textit{ie} the probability that the process never crosses $0$ during its $n$ first steps, and $F_{\underline{0},\infty}(k|x_0)$ is the probability of crossing 0 after exactly steps. We next make use of the  asymptotic behavior  of $q(x_0,n)$ obtained in \cite{Majumdar2017}, which yields  for $1\ll (x_0/a_\mu)^\mu \ll n$ : 
\begin{equation}\label{eq : survival_prob}
    q(x_0,n)\sim\frac{1}{\sqrt{n}}\frac{a_\mu^{-\frac{\mu}{2}}}{\sqrt{\pi} \Gamma(1+\frac{\mu}{2})} x_0^{\frac{\mu}{2}}.
\end{equation}
Combining  \eqref{eq : small_x0} and \eqref{eq : survival_prob} finally yields the  coefficient  $\alpha \sim n_{typ}/x^\mu$ defined above,  and thus  the following determination of    $n_{typ}$, valid for any $x_0\ll x$:
\begin{equation}\label{eq : result_n_typ}
    n_{typ}\sim\left[2^{\mu-1}\Gamma(\frac{1+\mu}{2})\right]^{-2}\Bigg(\frac{x}{a_\mu}\Bigg)^{\mu}.
\end{equation}
In order to determine the dependence on $x_0$ of the splitting probability, we use next the large $n$ behavior of the survival probability given by \cite{Majumdar2017}:
\begin{equation}\label{eq : survival_prob_2}
q(x_0,n)\underset{n\rightarrow\infty}{\sim}\frac{1}{\sqrt{n}}\left[\frac{1}{\sqrt{\pi}}+V(x_0)\right],   
\end{equation}
where  $V(x_0)$ is defined by its Laplace transform:
\begin{equation}\label{laplace}
\begin{split}
	\mathcal{L}V(\lambda)&= \int_0^\infty V(x_0)e^{-\lambda x_0}\text{d}x_0 \\ &=\frac{1}{\lambda\sqrt{\pi}}\left(\text{exp}\left[-\frac{\lambda}{\pi}\int_0^\infty \frac{\text{d}k}{\lambda^2+k^2}\text{ln}(1-\tilde{f}(k))\right]-1\right),
\end{split}
\end{equation}
and   $\tilde{f}(k)$  is the Fourier transformed jump distribution defined above.

Using equation \eqref{eq : time_partition_v2} and the above given asymptotic behavior of $n_{typ}$ \eqref{eq : result_n_typ}, we finally obtain the following general explicit  asymptotic determination of the splitting probability of jump processes :
\begin{equation}\label{aspi}
    \underset{x\rightarrow\infty}{\lim}\left[\frac{\pi_{0,\underline{x}}(x_0)}{A_\mu(x)} \right]=\frac{1}{\sqrt{\pi}}+V(x_0)
\end{equation}
where
\begin{equation}\label{amu}
    A_\mu(x)=\Bigg(\frac{a_\mu}{x}\Bigg)^{\mu/2} 2^{\mu-1}\Gamma\left(\frac{1+\mu}{2}\right).
\end{equation}
This  holds for {\it any} fixed $x_0$, including the regime $x_0\lesssim a_\mu$ that we intended to determine. This result thus elucidates the dependence of the splitting probability on $x$ (in the regime $x\gg x_0,a_\mu$), and, up to Laplace inversion, on $x_0$. In particular, the asymptotic behavior for  $x_0\ll a_\mu$ can be derived explicitly and yields:
\begin{equation}\label{eq : split_prob_uni}
    V(x_0)= \left\{
    \begin{array}{ll}
        -\left[\pi^{-\frac{3}{2}}\int_0^\infty\text{d}k\log(1-\Tilde{f}(k))\right]x_0+o(x_0) \\
        ~~~~~\text{if $\tilde{f}(k)\underset{k\rightarrow\infty}{=}o(k^{-1})$}\\
        \frac{\beta }{2\sqrt{\pi}\Gamma(1+\nu)\cos(\pi\nu/2)}x_0^\nu+o(x_0^\nu) \\ 
        ~~~~~\text{if $\tilde{f}(k)\underset{k\rightarrow\infty}{\sim} \beta k^{-\nu}$ with $\nu<1$}\\
            -\frac{\beta}{\pi^{3/2}}x_0\ln(x_0)+o\left(x_0\ln(x_0)\right)\\
        ~~~~~\text{if $\tilde{f}(k)\underset{k\rightarrow\infty}{\sim} \beta k^{-1}$}\\
    \end{array}
\right.
\end{equation}
Of note, the linear dependence of the auxiliary function $V(x_0)$ on $x_0$  obtained for  $\tilde{f}(k)\underset{k\rightarrow\infty}{=}o(k^{-1})$ in \eqref{eq : split_prob_uni} was given in \cite{Majumdar2017}. Interestingly, we find that the  scaling of the splitting probability with $x_0\ll a_\mu$ is not universal and can be sublinear depending solely on  the small scale behavior of the jump distribution $f(l)$; in particular it is independent of the large scale behavior of $f(l)$, and thus of $\mu$.

Remarkably, although $V(x_0)$ and thus $\pi_{0,\underline{x}}(x_0)$ (see \eqref{aspi})  generically depend on the jump process through the full jump distribution $f(l)$, the asymptotic transmission probability $\pi_{0,\underline{x}}(0)$ in fact depends on the jump distribution  only trough  $\mu$ and $a_\mu$  and takes the simple, explicit  form: 

\begin{equation}\label{eq : split_prob_uni_zero}
    \pi_{0,\underline{x}}(0)\underset{x\rightarrow\infty}{\sim}\frac{2^{\mu-1}}{\sqrt{\pi}}\Gamma\left(\frac{1+\mu}{2}\right)\Bigg(\frac{a_\mu}{x}
    \Bigg)^{\frac{\mu}{2}}.
\end{equation}
Even though the above derivation involves the uncontrolled asymptotics \eqref{eq : time_partition_v2}, we claim that our main  results \eqref{aspi} and \eqref{eq : split_prob_uni_zero} are exact; below we confirm these results either analytically  or numerically  on representative examples of jump processes.

{\it Jump processes with finite second moment.} 
We start by considering  continuous jump processes with a finite second moment, corresponding to the case $\mu=2$ in \eqref{fk}, which we illustrate by the class of Gamma jump processes of order $n>-1$, whose jump distributions read
\begin{equation}
f(l)=\frac{1}{2 \gamma^{n+1}\Gamma(n+1)}|l|^{n}e^{-\frac{|l|}{\gamma}},\label{gamma}
\end{equation} 
so that $a_2=\gamma\sqrt{(n+1)(n+2)/2}$. For $n=0$, this corresponds to the classical  exponential jump distribution 
 $f(l)= e^{-|l|/\gamma}/(2\gamma)$, for which, as mentioned above, the splitting probability is known exactly for all values of parameters \cite{Kampen:1992}, and satisfies in the regime   $x_0,a_2 \ll x$: 
%%\begin{equation}
%%\label{eq : split_laplace}
%%    \begin{split}
%%        &q(x_0,n)=\frac{1}{\sqrt{\pi n}}(1+\frac{x_0}{\gamma}) \\
%%        &\pi_{\underline{x},0}(x_0)=\frac{\gamma}{x}\left[1+\frac{x_0}{\gamma}\right]
%%    \end{split}
%%\end{equation}
\begin{equation}
\label{eq : split_laplace}
        \pi_{0,\underline{x}}(x_0)\underset{x\rightarrow\infty}{\sim}\frac{\gamma}{x}\left[1+\frac{x_0}{\gamma}\right].
\end{equation}
Calculating  $V(x_0)$ from \eqref{laplace}, one verifies explicitly the agreement of this exact result with \eqref{aspi}. Note that in this example $\tilde{f}(k)\underset{k\rightarrow\infty}{=}o(k^{-1})$, so that one verifies  in the $x_0\ll a_2 $ regime the linear dependence on $x_0$  predicted by \eqref{eq : split_prob_uni} (with the correct prefactor, see SM).

For $n=1$, one obtains the so--called Gamma jump process defined by the jump distribution $f(l)=\frac{1}{2\gamma^2}|l|e^{-|l|/\gamma}$. To the best of our knowledge the splitting probability for this jump process is not known; it can be obtained explicitly for all values of parameters as we proceed to show.
%%\begin{equation}
%%\label{eq : split_gamma}
%%    q(x_0,n)=\frac{1}{\sqrt{\pi n}}\left[\frac{2\sqrt{3}-1}{3}+\frac{x_0}{\gamma}+\frac{(\sqrt{3}-1)^2}{3}e^{-3x_0/\gamma}\right].
%%\end{equation}
 %Partitioning over the first step of the walk, we obtain the following equation for the splitting probability:
%\begin{equation}
%\label{eq : gamma_1}
%\pi_{\underline{x},0}(x_0)=\Bigg(\int_0^x f(y-x_0)\pi_{\underline{x},0}(y)\text{d}y+\int_x^\infty f(y-x_0)\text{d}y\Bigg)    
%\end{equation}
Let us denote by $D$ the differential operator, and check that  the following identity holds:
\begin{equation}
\label{eq : gamma_2}
    \left[D^2-\frac{1}{\gamma^2}\right]^2 f(y) = \frac{1}{\gamma^4}\delta(y)+\frac{1}{\gamma^2}\delta^{(2)}(y)
\end{equation}
where $\delta^{(2)}$ is the second derivative of the Dirac delta function. Applying the operator $(D^2-\gamma^{-2})^2$  to equation \eqref{int} yields (where derivatives are taken with respect to $x_0$) :
\begin{equation}
\label{eq : gamma_3}
\left[D^2-\frac{1}{\gamma^2}\right]^2\pi_{0,\underline{x}}(x_0)=\frac{1}{\gamma^4}\pi_{0,\underline{x}}(x_0)+\frac{1}{\gamma^2} D^2\pi_{0,\underline{x}}(x_0)
\end{equation}
and thus
\begin{equation}
    D^4 \pi_{0,\underline{x}}(x_0) - \frac{3}{\gamma^2} D^2\pi_{0,\underline{x}}(x_0)=0.
\end{equation}
The splitting probability is then obtained as:
\begin{equation}
\pi_{0,\underline{x}}(x_0)=Ae^{-\frac{\sqrt{3}}{\gamma} x_0}+Be^{+\frac{\sqrt{3}}{\gamma} x_0}+Cx_0+E
\end{equation}
where  $A,B,C,E$ are  determined by using \eqref{int}. This provides finally an explicit, exact determination of the splitting probability (see SM for explicit expressions) for all values of the parameters for the Gamma jump process. Calculating  $V(x_0)$ from \eqref{laplace}, one verifies explicitly the agreement of this exact result  for all $x_0\ll x$ with \eqref{aspi} (see SM).
In particular, in the $x_0 \ll a_2\ll x$ regime, the splitting probability satisfies:
\begin{equation}
\label{eq : gamma_4}
    \pi_{0,\underline{x}}(x_0)\underset{x\rightarrow\infty}{\sim}\frac{1}{x}\left[\sqrt{3}\gamma+(2\sqrt{3}-3)x_0+ o(x_0)\right].
\end{equation}
This linear scaling with $x_0$ is in agreement with equation \eqref{eq : split_prob_uni} (with the correct prefactor), as expected since $\tilde{f}(k)\underset{k\rightarrow\infty}{=}o(k^{-1})$.

\begin{figure}[h!]
    \centering
    \includegraphics[scale=0.46]{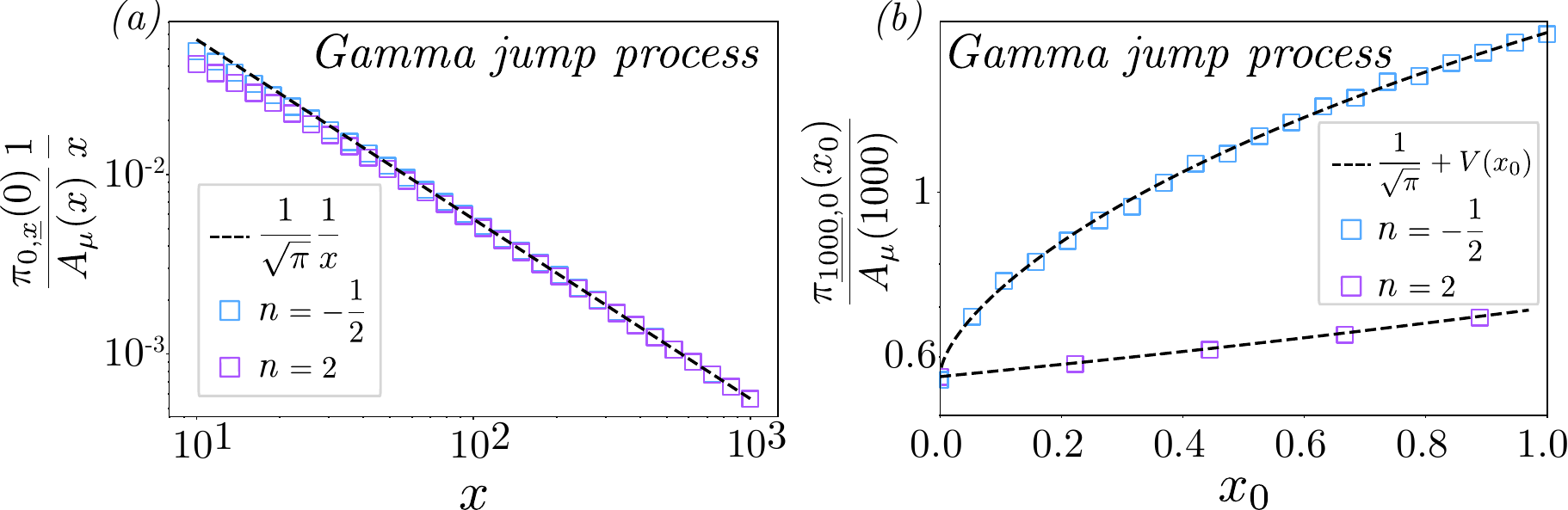}
    \caption{(a) Transmission probability for examples of Gamma jump processes. After rescaling according to \eqref{eq : split_prob_uni_zero}, the transmission probabilities collapse. (b) Small $x_0$ behavior of the splitting probability, as predicted by \eqref{aspi} and \eqref{eq : split_prob_uni}: for $n=-1/2$, one has $\nu=1/2$ and a sublinear dependence on $x_0$, while for $n=2$, one has a linear dependence on $x_0$. Theoretical predictions (dashed lines) are obtained by numerical inverse Laplace Transform of \eqref{laplace}, while simulations (squares) are averaged over $10^6$ trials.}
    \label{fig2}
\end{figure}

Finally, these two examples for $n=0,1$ provide analytical validations supporting the exactness of our results \eqref{aspi} and \eqref{eq : split_prob_uni_zero}. Additionally, we show in SM that the asymptotic splitting probability for higher or lower order  Gamma jump processes can be derived explicitly, and is confirmed by numerical simulations for $n=2$ and $n=-1/2$ in Fig. \ref{fig2}.

\begin{figure}[h!]
    \centering
    \includegraphics[scale=0.47]{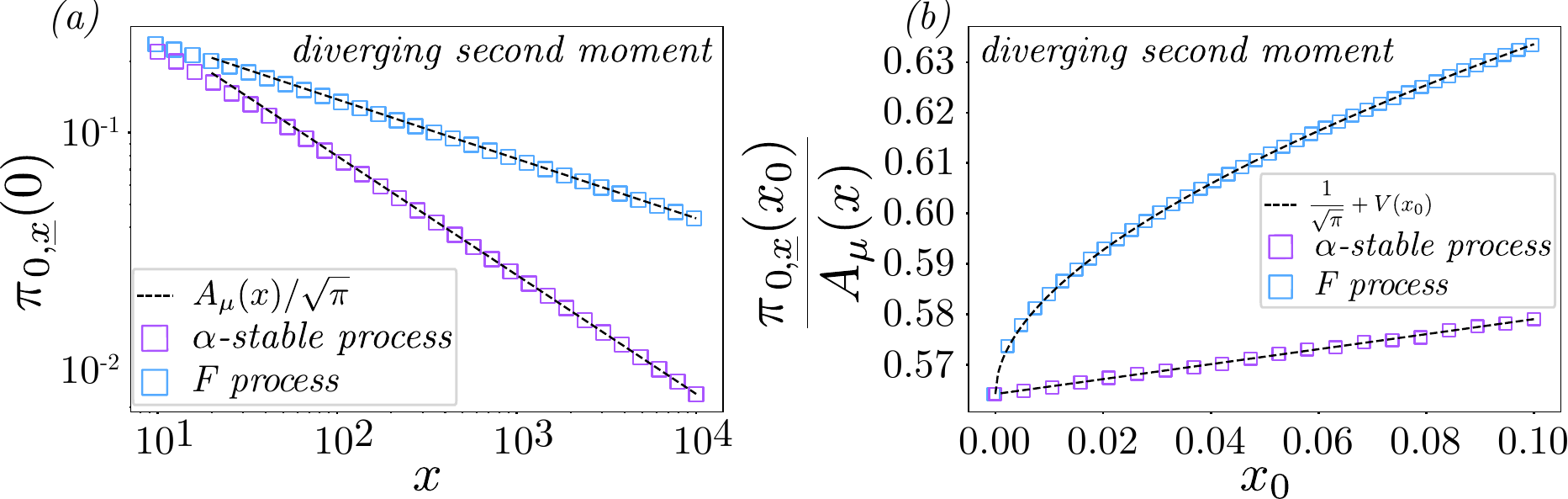}
    \caption{(a) Transmission probability for a  jump process with distribution $f(l)=(2\pi\sqrt{|l|}(1+l))^{-1}$ (denoted $F$ process), yielding $\mu=1/2$ and $\nu=1/2$ and a Levy flight with $\mu=1$ and $a_\mu=2$. The transmission probabilities (including prefactors) are accurately predicted. (b) Small $x_0$ behavior of the splitting probability. For the $F$ process, $x$ is fixed to $10^4$ and the behavior is sublinear. For the Levy Flight, $x$ is fixed to $2.10^5$ and one finds a  linear behavior. Theoretical predictions (dashed lines) are obtained by numerical inverse Laplace Transform of \eqref{laplace}, while simulations (squares) are averaged over $10^6$ trials.}
    \label{fig3}
\end{figure}

{\it Levy Flights.} For jump processes with infinite second moment \textit{ie} $\mu<2$ in \eqref{fk} -- called Levy flights \cite{Bouchaud:1990b,R.Metzler:2000,Zaburdaev:2015aa,Vezzani:2020aa}, no exact results for the splitting probability are available for generic $a_\mu,x_0$. We thus resort to numerical simulations to validate  predictions \eqref{aspi} to  \eqref{eq : split_prob_uni_zero} (see Fig. \ref{fig3}). First,  the prediction \eqref{eq : split_prob_uni_zero} of the transmission probability is confirmed and in particular fully  captures the dependence on $x$ (including the prefactor) that is controlled by the large scale behavior of $f(l)$, parameterized by $\mu$ and $a_\mu$ only.
In turn, \eqref{aspi} captures the dependence on $x_0$, which can lead to different scalings depending on the $l\to 0$  behavior of the jump distribution $f(l)$. The linear dependence on $x_0$ is illustrated  by the $\alpha$-stable jump distribution of parameter $\mu$ defined by $\tilde{f}(k)=e^{-(a_\mu |k|)^\mu}$, which verifies $\tilde{f}(k)\underset{k\rightarrow\infty}{=}o(k^{-1})$; an example of sublinear scaling with $x_0$ is provided by the jump distribution  $f(l)\propto \frac{1}{\sqrt{|l|}(1+|l|)}$, which corresponds to $\nu =1/2$ in \eqref{eq : split_prob_uni} and has an infinite second moment ($\mu=1/2$). Our results are thus also validated in the case of jump processes with infinite second moment.

\begin{figure}[h!]
    \centering
    \includegraphics[scale=0.48]{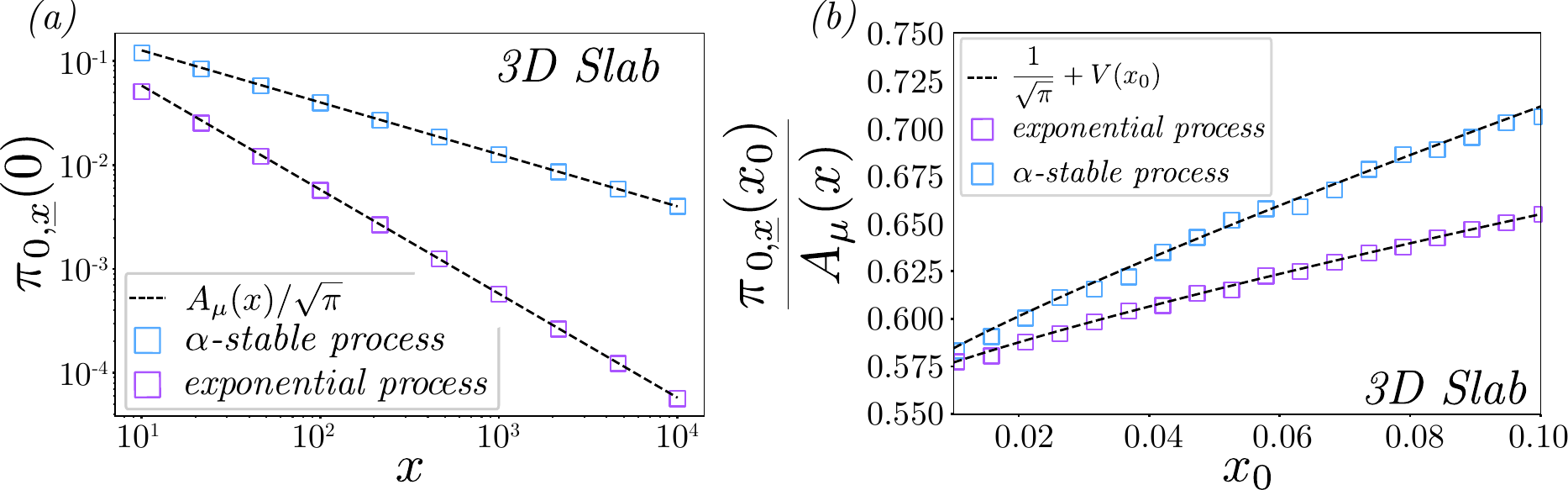}
    \caption{(a) Transmission probability for an exponential jump process with $\gamma=1$,  and a Levy flight with $\mu=1$ and $a_\mu=1$. The transmission probabilities (including prefactors)  are accurately predicted by \eqref{aspi} and \eqref{eq : split_prob_uni}. (b) Small $x_0$ behavior of the splitting probability. For both processes, $x$ is fixed to $10^3$ and the behavior is sublinear as predicted. Theoretical predictions (dashed  lines) are obtained by numerical inverse Laplace Transform of \eqref{laplace}, while simulations (squares) are averaged over $10^6$ trials.}
    \label{fig4}
\end{figure}

{\it Application to effective 1D problems.} In this section we show how our formalism applies to higher dimensional jump processes evolving between two parallel hyperplanes $H_1$ and $H_2$; coming back to our initial  example of the transmission of  particles (eg photons or neutrons)  through a slab of a scattering medium, the case $d=3$ is of particular interest. The trajectory is then naturally described as a 3$d$ jump process, where at each step, the direction of the jump is drawn  uniformly on the unit sphere and its length $r$ is drawn according to a distribution $p(r)$; typically experiments show that  exponential or Levy distributions $p(r)$ are observed, and provide as readout the transmission probability through the exit plane $H_2$ rather than $H_1$. Even if the problem is 3-dimensional, the determination of  the transmission probability amounts to solving for the splitting probability of a 1-dimensional problem, with the effective jump distribution $f(l)=\frac{1}{2}\int_{|l|}^\infty\frac{p(r)}{r} \text{d}r$  \cite{Mori2020a}. The above formalism  is thus directly applicable and provides explicit determinations of the asymptotic splitting probability and in particular of the transmission probability (see SM). In the case of an exponential jump distribution  $p(r)=\frac{1}{\gamma} e^{-\frac{r}{\gamma}}$, relevant to classical diffusive media \cite{Rotter:2017uy}, we obtain $f(l)=\frac{1}{2\gamma} \Gamma\left(0,\frac{|l|}{\gamma}   \right)$, where $\Gamma(x,y)$ stands for the incomplete Gamma function,  yielding $\tilde{f}(k)=\frac{\arctan(k\gamma)}{k\gamma}$ after Fourier transform. Equation \eqref{aspi} then provides -- up to Laplace inversion -- the asymptotic  expression (for $x\to\infty$)  of the splitting probability for any $x_0$. In particular   \eqref{aspi} and \eqref{eq : split_prob_uni}  yield for $x_0\ll a_2 \equiv \gamma/\sqrt{3}$:
\begin{equation}\label{eq : RTP_split}
    \pi_{0,\underline{x}}(x_0)\underset{x\rightarrow \infty}{\sim}\frac{1}{\sqrt{3}x}\left[\gamma-\frac{x_0\ln(x_0)}{2}+o\left(x_0\ln(x_0)\right)\right].
\end{equation}
In the case of $\alpha$-stable jump distributions, which have been shown recently to be relevant to photon scattering in hot atomic vapors \cite{PhysRevE.103.L010101,PhysRevE.90.052114}, we obtain $\tilde{f}(k)=\frac{\Gamma\left(\mu^{-1}\right)-\Gamma\left(\mu^{-1},(a_\mu k)^\mu\right)}{a_\mu \mu k}$ (see SM).   As above, this provides the asymptotic expression (for $x\to\infty$)  of the splitting probability for any $x_0$ thanks to \eqref{aspi},   and making use of   \eqref{eq : split_prob_uni}  one obtains for  $x_0\ll a_\mu$:
\begin{equation}\label{pi3d}
	\begin{split}
	    \pi_{0,\underline{x}}(x_0)\underset{x\rightarrow \infty}{\sim}&\frac{\Gamma\left(\frac{1+\mu}{2}\right)2^{\mu-1}}{\sqrt{(1+\mu)}}\left[\frac{a_\mu}{x}\right]^{\frac{\mu}{2}}\times\\
	    &\times\left[\frac{1}{\sqrt{\pi}}-\frac{\Gamma\left(\mu^{-1}\right)}{a_\mu \mu \pi^{\frac{3}{2}}}x_0\ln(x_0)(1+o(1))\right].
	\end{split}
\end{equation}
Agreement with simulations in both cases is displayed in Fig. \ref{fig4}.

{\it Conclusion.} We have  derived a  universal exact asymptotic form for the splitting probability for  continuous symmetric jump processes characterized by a  finite length scale $a_\mu$, which have proved to be relevant in various contexts, such as  transport of photons or neutrons  in scattering media.  This analysis covers the regime
  $x_0\ll a_\mu \ll x$ and  provides in particular a fully explicit determination of the transmission probability ($x_0=0$), in striking contrast with the trivial  prediction $ \pi_{0,\underline{x}}(x_0)=0$ obtained by taking  the continuous limit of the process.  This reveals  the importance of the microscopic properties of the dynamics. These results are illustrated with paradigmatic models of jump processes with applications to light scattering in heterogeneous media in realistic 3$d$ slab geometries. In this context, our explicit predictions of the transmission probability \eqref{pi3d}, which can be directly measured experimentally,  provides in principle a quantitative determination of not only the Levy exponent $\mu$, as already proposed in \cite{PhysRevE.103.L010101,PhysRevE.90.052114}, but also of the microscopic length scale $a_\mu$. This   significantly refines the characterization of the effective random process describing light scattering in the medium.

\bibliographystyle{apsrev4-2}
\bibliography{biblinew}

%apsrev4-2.bst 2019-01-14 (MD) hand-edited version of apsrev4-1.bst
%Control: key (0)
%Control: author (72) initials jnrlst
%Control: editor formatted (1) identically to author
%Control: production of article title (-1) disabled
%Control: page (0) single
%Control: year (1) truncated
%Control: production of eprint (0) enabled
\begin{thebibliography}{31}%
\makeatletter
\providecommand \@ifxundefined [1]{%
 \@ifx{#1\undefined}
}%
\providecommand \@ifnum [1]{%
 \ifnum #1\expandafter \@firstoftwo
 \else \expandafter \@secondoftwo
 \fi
}%
\providecommand \@ifx [1]{%
 \ifx #1\expandafter \@firstoftwo
 \else \expandafter \@secondoftwo
 \fi
}%
\providecommand \natexlab [1]{#1}%
\providecommand \enquote  [1]{``#1''}%
\providecommand \bibnamefont  [1]{#1}%
\providecommand \bibfnamefont [1]{#1}%
\providecommand \citenamefont [1]{#1}%
\providecommand \href@noop [0]{\@secondoftwo}%
\providecommand \href [0]{\begingroup \@sanitize@url \@href}%
\providecommand \@href[1]{\@@startlink{#1}\@@href}%
\providecommand \@@href[1]{\endgroup#1\@@endlink}%
\providecommand \@sanitize@url [0]{\catcode `\\12\catcode `\$12\catcode
  `\&12\catcode `\#12\catcode `\^12\catcode `\_12\catcode `\%12\relax}%
\providecommand \@@startlink[1]{}%
\providecommand \@@endlink[0]{}%
\providecommand \url  [0]{\begingroup\@sanitize@url \@url }%
\providecommand \@url [1]{\endgroup\@href {#1}{\urlprefix }}%
\providecommand \urlprefix  [0]{URL }%
\providecommand \Eprint [0]{\href }%
\providecommand \doibase [0]{https://doi.org/}%
\providecommand \selectlanguage [0]{\@gobble}%
\providecommand \bibinfo  [0]{\@secondoftwo}%
\providecommand \bibfield  [0]{\@secondoftwo}%
\providecommand \translation [1]{[#1]}%
\providecommand \BibitemOpen [0]{}%
\providecommand \bibitemStop [0]{}%
\providecommand \bibitemNoStop [0]{.\EOS\space}%
\providecommand \EOS [0]{\spacefactor3000\relax}%
\providecommand \BibitemShut  [1]{\csname bibitem#1\endcsname}%
\let\auto@bib@innerbib\@empty
%</preamble>
\bibitem [{\citenamefont {Redner}(2001)}]{Redner:2001a}%
  \BibitemOpen
  \bibfield  {author} {\bibinfo {author} {\bibfnamefont {S.}~\bibnamefont
  {Redner}},\ }\href@noop {} {\emph {\bibinfo {title} {A Guide to First-
  Passage Processes}}}\ (\bibinfo  {publisher} {Cambridge University Press,
  Cambridge, England},\ \bibinfo {year} {2001})\BibitemShut {NoStop}%
\bibitem [{\citenamefont {van Kampen}(1992)}]{Kampen:1992}%
  \BibitemOpen
  \bibfield  {author} {\bibinfo {author} {\bibfnamefont {N.~G.}\ \bibnamefont
  {van Kampen}},\ }\href@noop {} {\emph {\bibinfo {title} {Stochastic Processes
  in Physics and Chemistry}}}\ (\bibinfo  {publisher} {North-Holland,
  Amsterdam},\ \bibinfo {year} {1992})\BibitemShut {NoStop}%
\bibitem [{\citenamefont {Hughes}(1995)}]{Hughes:1995}%
  \BibitemOpen
  \bibfield  {author} {\bibinfo {author} {\bibfnamefont {B.}~\bibnamefont
  {Hughes}},\ }\href@noop {} {\emph {\bibinfo {title} {Random Walks and Random
  Environments}}}\ (\bibinfo  {publisher} {Oxford University Press, New York},\
  \bibinfo {year} {1995})\BibitemShut {NoStop}%
\bibitem [{\citenamefont {Gardiner}(2004)}]{Gardiner:2004}%
  \BibitemOpen
  \bibfield  {author} {\bibinfo {author} {\bibfnamefont {C.}~\bibnamefont
  {Gardiner}},\ }\href@noop {} {\emph {\bibinfo {title} {Handbook of Stochastic
  Methods for Physics, Chemistry and Natural Sciences}}}\ (\bibinfo
  {publisher} {Springer},\ \bibinfo {year} {2004})\BibitemShut {NoStop}%
\bibitem [{\citenamefont {Condamin}\ \emph {et~al.}(2008)\citenamefont
  {Condamin}, \citenamefont {Tejedor}, \citenamefont {Voituriez}, \citenamefont
  {Benichou},\ and\ \citenamefont {Klafter}}]{Condamin:2008}%
  \BibitemOpen
  \bibfield  {author} {\bibinfo {author} {\bibfnamefont {S.}~\bibnamefont
  {Condamin}}, \bibinfo {author} {\bibfnamefont {V.}~\bibnamefont {Tejedor}},
  \bibinfo {author} {\bibfnamefont {R.}~\bibnamefont {Voituriez}}, \bibinfo
  {author} {\bibfnamefont {O.}~\bibnamefont {Benichou}},\ and\ \bibinfo
  {author} {\bibfnamefont {J.}~\bibnamefont {Klafter}},\ }\href
  {http://www.pnas.org/cgi/content/abstract/0712158105v1} {\bibfield  {journal}
  {\bibinfo  {journal} {Proceedings of the National Academy of Sciences}\
  }\textbf {\bibinfo {volume} {105}},\ \bibinfo {pages} {5675} (\bibinfo {year}
  {2008})}\BibitemShut {NoStop}%
\bibitem [{\citenamefont {Chevalier}\ \emph {et~al.}(2011)\citenamefont
  {Chevalier}, \citenamefont {B\'enichou}, \citenamefont {Meyer},\ and\
  \citenamefont {Voituriez}}]{al:2011}%
  \BibitemOpen
  \bibfield  {author} {\bibinfo {author} {\bibfnamefont {C.}~\bibnamefont
  {Chevalier}}, \bibinfo {author} {\bibfnamefont {O.}~\bibnamefont
  {B\'enichou}}, \bibinfo {author} {\bibfnamefont {B.}~\bibnamefont {Meyer}},\
  and\ \bibinfo {author} {\bibfnamefont {R.}~\bibnamefont {Voituriez}},\ }\href
  {http://stacks.iop.org/1751-8121/44/i=2/a=025002} {\bibfield  {journal}
  {\bibinfo  {journal} {Journal of Physics A: Mathematical and Theoretical}\
  }\textbf {\bibinfo {volume} {44}},\ \bibinfo {pages} {025002} (\bibinfo
  {year} {2011})}\BibitemShut {NoStop}%
\bibitem [{\citenamefont {Moran}(1962)}]{moran}%
  \BibitemOpen
  \bibfield  {author} {\bibinfo {author} {\bibfnamefont {P.}~\bibnamefont
  {Moran}},\ }\href@noop {} {\emph {\bibinfo {title} {The Statistical processes
  of of evolutionary theory}}}\ (\bibinfo  {publisher} {Oxford University
  Press},\ \bibinfo {year} {1962})\BibitemShut {NoStop}%
\bibitem [{\citenamefont {Oshanin}\ and\ \citenamefont
  {Redner}(2009)}]{Oshanin_2009}%
  \BibitemOpen
  \bibfield  {author} {\bibinfo {author} {\bibfnamefont {G.}~\bibnamefont
  {Oshanin}}\ and\ \bibinfo {author} {\bibfnamefont {S.}~\bibnamefont
  {Redner}},\ }\href {https://doi.org/10.1209/0295-5075/85/10008} {\bibfield
  {journal} {\bibinfo  {journal} {{EPL} (Europhysics Letters)}\ }\textbf
  {\bibinfo {volume} {85}},\ \bibinfo {pages} {10008} (\bibinfo {year}
  {2009})}\BibitemShut {NoStop}%
\bibitem [{\citenamefont {Rotter}\ and\ \citenamefont
  {Gigan}(2017)}]{Rotter:2017uy}%
  \BibitemOpen
  \bibfield  {author} {\bibinfo {author} {\bibfnamefont {S.}~\bibnamefont
  {Rotter}}\ and\ \bibinfo {author} {\bibfnamefont {S.}~\bibnamefont {Gigan}},\
  }\href {https://doi.org/10.1103/RevModPhys.89.015005} {\bibfield  {journal}
  {\bibinfo  {journal} {Reviews of Modern Physics}\ }\textbf {\bibinfo {volume}
  {89}},\ \bibinfo {pages} {015005} (\bibinfo {year} {2017})}\BibitemShut
  {NoStop}%
\bibitem [{\citenamefont {Burioni}\ \emph {et~al.}(2010)\citenamefont
  {Burioni}, \citenamefont {Caniparoli},\ and\ \citenamefont
  {Vezzani}}]{Burioni:2010tr}%
  \BibitemOpen
  \bibfield  {author} {\bibinfo {author} {\bibfnamefont {R.}~\bibnamefont
  {Burioni}}, \bibinfo {author} {\bibfnamefont {L.}~\bibnamefont
  {Caniparoli}},\ and\ \bibinfo {author} {\bibfnamefont {A.}~\bibnamefont
  {Vezzani}},\ }\href {https://doi.org/10.1103/PhysRevE.81.060101} {\bibfield
  {journal} {\bibinfo  {journal} {Physical Review E}\ }\textbf {\bibinfo
  {volume} {81}},\ \bibinfo {pages} {060101} (\bibinfo {year}
  {2010})}\BibitemShut {NoStop}%
\bibitem [{\citenamefont {Burioni}\ \emph {et~al.}(2014)\citenamefont
  {Burioni}, \citenamefont {Ubaldi},\ and\ \citenamefont
  {Vezzani}}]{PhysRevE.89.022135}%
  \BibitemOpen
  \bibfield  {author} {\bibinfo {author} {\bibfnamefont {R.}~\bibnamefont
  {Burioni}}, \bibinfo {author} {\bibfnamefont {E.}~\bibnamefont {Ubaldi}},\
  and\ \bibinfo {author} {\bibfnamefont {A.}~\bibnamefont {Vezzani}},\ }\href
  {https://doi.org/10.1103/PhysRevE.89.022135} {\bibfield  {journal} {\bibinfo
  {journal} {Phys. Rev. E}\ }\textbf {\bibinfo {volume} {89}},\ \bibinfo
  {pages} {022135} (\bibinfo {year} {2014})}\BibitemShut {NoStop}%
\bibitem [{\citenamefont {Baudouin}\ \emph {et~al.}(2014)\citenamefont
  {Baudouin}, \citenamefont {Pierrat}, \citenamefont {Eloy}, \citenamefont
  {Nunes-Pereira}, \citenamefont {Cuniasse}, \citenamefont {Mercadier},\ and\
  \citenamefont {Kaiser}}]{PhysRevE.90.052114}%
  \BibitemOpen
  \bibfield  {author} {\bibinfo {author} {\bibfnamefont {Q.}~\bibnamefont
  {Baudouin}}, \bibinfo {author} {\bibfnamefont {R.}~\bibnamefont {Pierrat}},
  \bibinfo {author} {\bibfnamefont {A.}~\bibnamefont {Eloy}}, \bibinfo {author}
  {\bibfnamefont {E.~J.}\ \bibnamefont {Nunes-Pereira}}, \bibinfo {author}
  {\bibfnamefont {P.-A.}\ \bibnamefont {Cuniasse}}, \bibinfo {author}
  {\bibfnamefont {N.}~\bibnamefont {Mercadier}},\ and\ \bibinfo {author}
  {\bibfnamefont {R.}~\bibnamefont {Kaiser}},\ }\href
  {https://doi.org/10.1103/PhysRevE.90.052114} {\bibfield  {journal} {\bibinfo
  {journal} {Phys. Rev. E}\ }\textbf {\bibinfo {volume} {90}},\ \bibinfo
  {pages} {052114} (\bibinfo {year} {2014})}\BibitemShut {NoStop}%
\bibitem [{\citenamefont {Ara\'ujo}\ \emph {et~al.}(2021)\citenamefont
  {Ara\'ujo}, \citenamefont {de~Silans},\ and\ \citenamefont
  {Kaiser}}]{PhysRevE.103.L010101}%
  \BibitemOpen
  \bibfield  {author} {\bibinfo {author} {\bibfnamefont {M.~O.}\ \bibnamefont
  {Ara\'ujo}}, \bibinfo {author} {\bibfnamefont {T.~P.}\ \bibnamefont
  {de~Silans}},\ and\ \bibinfo {author} {\bibfnamefont {R.}~\bibnamefont
  {Kaiser}},\ }\href {https://doi.org/10.1103/PhysRevE.103.L010101} {\bibfield
  {journal} {\bibinfo  {journal} {Phys. Rev. E}\ }\textbf {\bibinfo {volume}
  {103}},\ \bibinfo {pages} {L010101} (\bibinfo {year} {2021})}\BibitemShut
  {NoStop}%
\bibitem [{\citenamefont {Levernier}\ \emph {et~al.}(2021)\citenamefont
  {Levernier}, \citenamefont {B{\'e}nichou},\ and\ \citenamefont
  {Voituriez}}]{Levernier:2021aa}%
  \BibitemOpen
  \bibfield  {author} {\bibinfo {author} {\bibfnamefont {N.}~\bibnamefont
  {Levernier}}, \bibinfo {author} {\bibfnamefont {O.}~\bibnamefont
  {B{\'e}nichou}},\ and\ \bibinfo {author} {\bibfnamefont {R.}~\bibnamefont
  {Voituriez}},\ }\href {https://doi.org/10.1103/PhysRevLett.126.100602}
  {\bibfield  {journal} {\bibinfo  {journal} {Physical Review Letters}\
  }\textbf {\bibinfo {volume} {126}},\ \bibinfo {pages} {100602} (\bibinfo
  {year} {2021})}\BibitemShut {NoStop}%
\bibitem [{\citenamefont {Majumdar}(2010)}]{MAJUMDAR20104299}%
  \BibitemOpen
  \bibfield  {author} {\bibinfo {author} {\bibfnamefont {S.~N.}\ \bibnamefont
  {Majumdar}},\ }\href
  {https://doi.org/https://doi.org/10.1016/j.physa.2010.01.021} {\bibfield
  {journal} {\bibinfo  {journal} {Physica A: Statistical Mechanics and its
  Applications}\ }\textbf {\bibinfo {volume} {389}},\ \bibinfo {pages} {4299}
  (\bibinfo {year} {2010})},\ \bibinfo {note} {proceedings of the 12th
  International Summer School on Fundamental Problems in Statistical
  Physics}\BibitemShut {NoStop}%
\bibitem [{\citenamefont {Majumdar}\ \emph {et~al.}(2006)\citenamefont
  {Majumdar}, \citenamefont {Comtet},\ and\ \citenamefont
  {Ziff}}]{Majumdar:2006aa}%
  \BibitemOpen
  \bibfield  {author} {\bibinfo {author} {\bibfnamefont {S.~N.}\ \bibnamefont
  {Majumdar}}, \bibinfo {author} {\bibfnamefont {A.}~\bibnamefont {Comtet}},\
  and\ \bibinfo {author} {\bibfnamefont {R.~M.}\ \bibnamefont {Ziff}},\ }\href
  {https://doi.org/10.1007/s10955-005-9002-x} {\bibfield  {journal} {\bibinfo
  {journal} {Journal of Statistical Physics}\ }\textbf {\bibinfo {volume}
  {122}},\ \bibinfo {pages} {833} (\bibinfo {year} {2006})}\BibitemShut
  {NoStop}%
\bibitem [{\citenamefont {Bouchaud}\ and\ \citenamefont
  {Georges}(1990)}]{Bouchaud:1990b}%
  \BibitemOpen
  \bibfield  {author} {\bibinfo {author} {\bibfnamefont {J.-P.}\ \bibnamefont
  {Bouchaud}}\ and\ \bibinfo {author} {\bibfnamefont {A.}~\bibnamefont
  {Georges}},\ }\href
  {http://www.sciencedirect.com/science/article/B6TVP-46SXPMN-7F/2/691b5d0049eff14ec4a52e098fae92db}
  {\bibfield  {journal} {\bibinfo  {journal} {Physics Reports}\ }\textbf
  {\bibinfo {volume} {195}},\ \bibinfo {pages} {127} (\bibinfo {year}
  {1990})}\BibitemShut {NoStop}%
\bibitem [{\citenamefont {R.Metzler}\ and\ \citenamefont
  {J.Klafter}(2000)}]{R.Metzler:2000}%
  \BibitemOpen
  \bibfield  {author} {\bibinfo {author} {\bibnamefont {R.Metzler}}\ and\
  \bibinfo {author} {\bibnamefont {J.Klafter}},\ }\href@noop {} {\bibfield
  {journal} {\bibinfo  {journal} {Phys. Rep.}\ }\textbf {\bibinfo {volume}
  {339}},\ \bibinfo {pages} {1} (\bibinfo {year} {2000})}\BibitemShut {NoStop}%
\bibitem [{\citenamefont {Widom}(1961)}]{Widom1961}%
  \BibitemOpen
  \bibfield  {author} {\bibinfo {author} {\bibfnamefont {H.}~\bibnamefont
  {Widom}},\ }\href {https://doi.org/10.2307/1993340} {\bibfield  {journal}
  {\bibinfo  {journal} {Transactions of the American Mathematical Society}\
  }\textbf {\bibinfo {volume} {98}},\ \bibinfo {pages} {430} (\bibinfo {year}
  {1961})}\BibitemShut {NoStop}%
\bibitem [{\citenamefont {Blumenthal}\ \emph {et~al.}(1961)\citenamefont
  {Blumenthal}, \citenamefont {Getoor},\ and\ \citenamefont
  {Ray}}]{Blumenthal1961}%
  \BibitemOpen
  \bibfield  {author} {\bibinfo {author} {\bibfnamefont {R.~M.}\ \bibnamefont
  {Blumenthal}}, \bibinfo {author} {\bibfnamefont {R.~K.}\ \bibnamefont
  {Getoor}},\ and\ \bibinfo {author} {\bibfnamefont {D.~B.}\ \bibnamefont
  {Ray}},\ }\href {https://doi.org/10.1090/s0002-9947-1961-0126885-4}
  {\bibfield  {journal} {\bibinfo  {journal} {Transactions of the American
  Mathematical Society}\ }\textbf {\bibinfo {volume} {99}},\ \bibinfo {pages}
  {540} (\bibinfo {year} {1961})}\BibitemShut {NoStop}%
\bibitem [{\citenamefont {Majumdar}\ \emph {et~al.}(2010)\citenamefont
  {Majumdar}, \citenamefont {Rosso},\ and\ \citenamefont
  {Zoia}}]{Majumdar:2010}%
  \BibitemOpen
  \bibfield  {author} {\bibinfo {author} {\bibfnamefont {S.~N.}\ \bibnamefont
  {Majumdar}}, \bibinfo {author} {\bibfnamefont {A.}~\bibnamefont {Rosso}},\
  and\ \bibinfo {author} {\bibfnamefont {A.}~\bibnamefont {Zoia}},\ }\href
  {http://link.aps.org/abstract/PRL/v104/e020602} {\bibfield  {journal}
  {\bibinfo  {journal} {Physical Review Letters}\ }\textbf {\bibinfo {volume}
  {104}} (\bibinfo {year} {2010})}\BibitemShut {NoStop}%
\bibitem [{Note1()}]{Note1}%
  \BibitemOpen
  \bibinfo {note} {Remarkably, this scaling behavior with $x_0/x$ has been
  generalized to non Markovian scale invariant processes, at the cost of
  determining the persistence exponent $\theta $.}\BibitemShut {Stop}%
\bibitem [{\citenamefont {Ziff}\ \emph {et~al.}(2009)\citenamefont {Ziff},
  \citenamefont {Majumdar},\ and\ \citenamefont {Comtet}}]{Ziff:2009vh}%
  \BibitemOpen
  \bibfield  {author} {\bibinfo {author} {\bibfnamefont {R.~M.}\ \bibnamefont
  {Ziff}}, \bibinfo {author} {\bibfnamefont {S.~N.}\ \bibnamefont {Majumdar}},\
  and\ \bibinfo {author} {\bibfnamefont {A.}~\bibnamefont {Comtet}},\ }\href
  {https://doi.org/10.1063/1.3137062} {\bibfield  {journal} {\bibinfo
  {journal} {The Journal of Chemical Physics}\ }\textbf {\bibinfo {volume}
  {130}},\ \bibinfo {pages} {204104} (\bibinfo {year} {2009})}\BibitemShut
  {NoStop}%
\bibitem [{\citenamefont {Romanczuk}\ \emph {et~al.}(2012)\citenamefont
  {Romanczuk}, \citenamefont {Bar}, \citenamefont {Ebeling}, \citenamefont
  {Lindner},\ and\ \citenamefont {Schimansky-Geier}}]{Romanczuk:2012fk}%
  \BibitemOpen
  \bibfield  {author} {\bibinfo {author} {\bibfnamefont {P.}~\bibnamefont
  {Romanczuk}}, \bibinfo {author} {\bibfnamefont {M.}~\bibnamefont {Bar}},
  \bibinfo {author} {\bibfnamefont {W.}~\bibnamefont {Ebeling}}, \bibinfo
  {author} {\bibfnamefont {B.}~\bibnamefont {Lindner}},\ and\ \bibinfo {author}
  {\bibfnamefont {L.}~\bibnamefont {Schimansky-Geier}},\ }\href
  {https://doi.org/10.1140/epjst/e2012-01529-y} {\bibfield  {journal} {\bibinfo
   {journal} {EPJE-ST}\ }\textbf {\bibinfo {volume} {202}},\ \bibinfo {pages}
  {1} (\bibinfo {year} {2012})}\BibitemShut {NoStop}%
\bibitem [{\citenamefont {Tejedor}\ \emph {et~al.}(2012)\citenamefont
  {Tejedor}, \citenamefont {Voituriez},\ and\ \citenamefont
  {B{\'e}nichou}}]{Tejedor:2012ly}%
  \BibitemOpen
  \bibfield  {author} {\bibinfo {author} {\bibfnamefont {V.}~\bibnamefont
  {Tejedor}}, \bibinfo {author} {\bibfnamefont {R.}~\bibnamefont {Voituriez}},\
  and\ \bibinfo {author} {\bibfnamefont {O.}~\bibnamefont {B{\'e}nichou}},\
  }\href {http://link.aps.org/doi/10.1103/PhysRevLett.108.088103} {\bibfield
  {journal} {\bibinfo  {journal} {Physical Review Letters}\ }\textbf {\bibinfo
  {volume} {108}},\ \bibinfo {pages} {088103} (\bibinfo {year}
  {2012})}\BibitemShut {NoStop}%
\bibitem [{\citenamefont {Meyer}\ and\ \citenamefont
  {Rieger}(2021)}]{Meyer:2021tx}%
  \BibitemOpen
  \bibfield  {author} {\bibinfo {author} {\bibfnamefont {H.}~\bibnamefont
  {Meyer}}\ and\ \bibinfo {author} {\bibfnamefont {H.}~\bibnamefont {Rieger}},\
  }\href {https://doi.org/10.1103/PhysRevLett.127.070601} {\bibfield  {journal}
  {\bibinfo  {journal} {Physical Review Letters}\ }\textbf {\bibinfo {volume}
  {127}},\ \bibinfo {pages} {070601} (\bibinfo {year} {2021})}\BibitemShut
  {NoStop}%
\bibitem [{\citenamefont {Mori}\ \emph {et~al.}(2020)\citenamefont {Mori},
  \citenamefont {{Le Doussal}}, \citenamefont {Majumdar},\ and\ \citenamefont
  {Schehr}}]{Mori2020a}%
  \BibitemOpen
  \bibfield  {author} {\bibinfo {author} {\bibfnamefont {F.}~\bibnamefont
  {Mori}}, \bibinfo {author} {\bibfnamefont {P.}~\bibnamefont {{Le Doussal}}},
  \bibinfo {author} {\bibfnamefont {S.~N.}\ \bibnamefont {Majumdar}},\ and\
  \bibinfo {author} {\bibfnamefont {G.}~\bibnamefont {Schehr}},\ }\href
  {https://doi.org/10.1103/PhysRevLett.124.090603} {\bibfield  {journal}
  {\bibinfo  {journal} {Physical Review Letters}\ }\textbf {\bibinfo {volume}
  {124}},\ \bibinfo {pages} {1} (\bibinfo {year} {2020})},\ \Eprint
  {https://arxiv.org/abs/2001.01492} {arXiv:2001.01492} \BibitemShut {NoStop}%
\bibitem [{\citenamefont {Levernier}\ \emph {et~al.}(2018)\citenamefont
  {Levernier}, \citenamefont {B{\'e}nichou}, \citenamefont {Gu{\'e}rin},\ and\
  \citenamefont {Voituriez}}]{Levernier:2018qf}%
  \BibitemOpen
  \bibfield  {author} {\bibinfo {author} {\bibfnamefont {N.}~\bibnamefont
  {Levernier}}, \bibinfo {author} {\bibfnamefont {O.}~\bibnamefont
  {B{\'e}nichou}}, \bibinfo {author} {\bibfnamefont {T.}~\bibnamefont
  {Gu{\'e}rin}},\ and\ \bibinfo {author} {\bibfnamefont {R.}~\bibnamefont
  {Voituriez}},\ }\href {https://doi.org/10.1103/PhysRevE.98.022125} {\bibfield
   {journal} {\bibinfo  {journal} {Physical Review E}\ }\textbf {\bibinfo
  {volume} {98}},\ \bibinfo {pages} {022125} (\bibinfo {year}
  {2018})}\BibitemShut {NoStop}%
\bibitem [{\citenamefont {Majumdar}\ \emph {et~al.}(2017)\citenamefont
  {Majumdar}, \citenamefont {Mounaix},\ and\ \citenamefont
  {Schehr}}]{Majumdar2017}%
  \BibitemOpen
  \bibfield  {author} {\bibinfo {author} {\bibfnamefont {S.~N.}\ \bibnamefont
  {Majumdar}}, \bibinfo {author} {\bibfnamefont {P.}~\bibnamefont {Mounaix}},\
  and\ \bibinfo {author} {\bibfnamefont {G.}~\bibnamefont {Schehr}},\ }\href
  {https://doi.org/10.1088/1751-8121/aa8d28} {\bibfield  {journal} {\bibinfo
  {journal} {Journal of Physics A: Mathematical and Theoretical}\ }\textbf
  {\bibinfo {volume} {50}},\ \bibinfo {pages} {465002} (\bibinfo {year}
  {2017})}\BibitemShut {NoStop}%
\bibitem [{\citenamefont {Zaburdaev}\ \emph {et~al.}(2015)\citenamefont
  {Zaburdaev}, \citenamefont {Denisov},\ and\ \citenamefont
  {Klafter}}]{Zaburdaev:2015aa}%
  \BibitemOpen
  \bibfield  {author} {\bibinfo {author} {\bibfnamefont {V.}~\bibnamefont
  {Zaburdaev}}, \bibinfo {author} {\bibfnamefont {S.}~\bibnamefont {Denisov}},\
  and\ \bibinfo {author} {\bibfnamefont {J.}~\bibnamefont {Klafter}},\ }\href
  {https://doi.org/10.1103/RevModPhys.87.483} {\bibfield  {journal} {\bibinfo
  {journal} {Reviews of Modern Physics}\ }\textbf {\bibinfo {volume} {87}},\
  \bibinfo {pages} {483} (\bibinfo {year} {2015})}\BibitemShut {NoStop}%
\bibitem [{\citenamefont {Vezzani}\ \emph {et~al.}(2020)\citenamefont
  {Vezzani}, \citenamefont {Barkai},\ and\ \citenamefont
  {Burioni}}]{Vezzani:2020aa}%
  \BibitemOpen
  \bibfield  {author} {\bibinfo {author} {\bibfnamefont {A.}~\bibnamefont
  {Vezzani}}, \bibinfo {author} {\bibfnamefont {E.}~\bibnamefont {Barkai}},\
  and\ \bibinfo {author} {\bibfnamefont {R.}~\bibnamefont {Burioni}},\ }\href
  {https://doi.org/10.1038/s41598-020-59187-w} {\bibfield  {journal} {\bibinfo
  {journal} {Scientific Reports}\ }\textbf {\bibinfo {volume} {10}},\ \bibinfo
  {pages} {2732} (\bibinfo {year} {2020})}\BibitemShut {NoStop}%
\end{thebibliography}%



\section*{Supplementary material}

# Supplementary Material for "Splitting Probabilities of Jump Processes"

J. Klinger and O. Bénichou
*Laboratoire de Physique Théorique de la Matière Condensée,*
*CNRS, UPMC, 4 Place Jussieu, 75005 Paris, France*

R. Voituriez
*Laboratoire Jean Perrin, CNRS, UPMC, 4 Place Jussieu, 75005 Paris, France*
(Dated: January 31, 2022)

This Supplementary Material presents details of calculations supporting the main text. We provide :

- details on the asymptotic behavior of $V(x_0)$ in the $x_0 \ll a_\mu$ limit.
- details on the derivation of the splitting probability for Laplace jump processes.
- details on the derivation of the splitting probability for Gamma jump processes.
- details on the Fourier Transforms of the Gamma jump process family.
- details on the derivation of effective jump distributions in $3d$ geometries.

## I. SMALL $x_0$ BEHAVIOR OF $V(x_0)$

In this section we provide the derivation of the asymptotic behavior of $V(x_0)$, defined by its Laplace transform:

$$\mathcal{L}V(\lambda) = \int_0^\infty V(x_0) e^{-\lambda x_0} \mathrm{d}x_0$$
$$= \frac{1}{\lambda\sqrt{\pi}} \left( \exp\left[ -\frac{\lambda}{\pi} \int_0^\infty \frac{\mathrm{d}k}{\lambda^2 + k^2} \ln(1-\tilde{f}(k)) \right] - 1 \right), \tag{S1}$$

with $\tilde{f}(k)$ the Fourier transform of the jump distribution $f(l)$. The small $x_0$ behavior is given by the corresponding large $\lambda$ behavior of the Laplace Transform.

### A. $\tilde{f}(k) \underset{k\to\infty}{=} o(k^{-1})$ case

Note that this case has already been addressed in[2]. However, we reproduce the derivation for completeness. For $\tilde{f}(k) \underset{k\to\infty}{=} o(k^{-1})$, the integral $\int_0^\infty \ln(1-\tilde{f}(k))\mathrm{d}k$ converges and, to first order, one has:

$$\exp\left[ -\frac{\lambda}{\pi} \int_0^\infty \frac{\mathrm{d}k}{\lambda^2 + k^2} \ln(1-\tilde{f}(k)) \right] = 1 - \frac{1}{\lambda\pi}\int_0^\infty \ln(1-\tilde{f}(k))\mathrm{d}k + o\left(\frac{1}{\lambda}\right) \tag{S2}$$

yielding:

$$\mathcal{L}V(\lambda) \underset{\lambda\to\infty}{=} -\frac{1}{\lambda^2 \pi^{\frac{3}{2}}} \int_0^\infty \ln(1-\tilde{f}(k))\mathrm{d}k + o\left(\frac{1}{\lambda^2}\right) \tag{S3}$$

and thus the corresponding $x_0$ behavior:

$$V(x_0) \underset{x_0\to 0}{=} -\frac{x_0}{\pi^{\frac{3}{2}}} \int_0^\infty \ln(1-\tilde{f}(k))\mathrm{d}k + o(x_0) \tag{S4}$$

**B.** $\tilde{f}(k) \underset{k\to\infty}{\sim} \beta k^{-\nu}$ **with $\nu < 1$ case**

In this case, the integral $\int_0^\infty \ln(1 - \tilde{f}(k))dk$ is ill-defined. We thus need to take further care in making the large $\lambda$ expansion. Let us denote $A(\lambda) = \mathcal{L}V(\lambda) + \frac{1}{\lambda\sqrt{\pi}}$. Performing the change of variable $\frac{k}{\lambda} = u$, one obtains:

$$A(\lambda) = \frac{1}{\lambda\sqrt{\pi}}\exp\left[-\frac{\lambda}{\pi}\int_0^\infty \frac{dk}{\lambda^2 + k^2} \ln\left(1 - \tilde{f}(k)\right)\right]$$
$$= \frac{1}{\lambda\sqrt{\pi}}\exp\left[-\frac{1}{\pi}\int_0^\infty \frac{du}{1 + u^2} \ln\left(1 - \tilde{f}(\lambda u)\right)\right]$$

and using the large $k$ behavior of $\tilde{f}(k)$:

$$\begin{aligned}A(\lambda) &= \frac{1}{\lambda\sqrt{\pi}}\exp\left[\frac{\beta}{\lambda^\nu}\frac{1}{\pi}\int_0^\infty \frac{du}{1+u^2}\frac{1}{u^\nu} + o(\frac{1}{\lambda^\nu})\right]\\ &= \frac{1}{\lambda\sqrt{\pi}}\exp\left[\frac{\beta}{\lambda^\nu}\frac{1}{2\cos(\pi\nu/2)} + o(\frac{1}{\lambda^\nu})\right]\\ &= \frac{1}{\lambda\sqrt{\pi}}\left[1 + \frac{\beta}{\lambda^\nu}\frac{1}{2\cos(\pi\nu/2)} + o(\frac{1}{\lambda^\nu})\right]\\ &= \frac{1}{\lambda\sqrt{\pi}} + \frac{\beta}{\lambda^{1+\nu}}\frac{1}{2\cos(\pi\nu/2)\sqrt{\pi}} + o(\frac{1}{\lambda^{1+\nu}})\end{aligned}$$ (S5)

Finally, the inverse Laplace Transform yields the following small $x_0$ behavior :

$$V(x_0) \underset{x_0 \to 0}{=} \frac{\beta}{2\Gamma(1+\nu)\cos(\pi\nu/2)\sqrt{\pi}} x_0^\nu + o(x_0^\nu)$$ (S6)

**C.** $\tilde{f}(k) \underset{k\to\infty}{\sim} \beta k^{-1}$ **case**

In this case, the change of variable used above cannot be directly applied since the integral $\int_0^\infty \frac{du}{1+u^2}\frac{1}{u}$ is ill-defined. Let us denote $\tilde{F}(k)$ a primitive of $\ln(1 - \tilde{f}(k))$. Its large $k$ behavior is dictated by that of $\tilde{f}$ and reads : $\tilde{F}(k) \sim -\beta \ln(k)$. Let us now extract the dominating large $\lambda$ behavior of $A(\lambda)$ :

$$\begin{aligned}A(\lambda) &= \frac{1}{\lambda\sqrt{\pi}}\exp\left[-\frac{1}{\pi}\int_0^\infty \frac{du}{1+u^2}\ln\left(1 - \tilde{f}(\lambda u)\right)\right]\\ &= \frac{1}{\lambda\sqrt{\pi}}\exp\left[-\left[\frac{\tilde{F}(\lambda u)}{\pi\lambda(1+u^2)}\right]_0^\infty - \frac{1}{\pi\lambda}\int_0^\infty \frac{2udu}{(1+u^2)^2}\tilde{F}(\lambda u)\right]\\ &\underset{\lambda\to\infty}{=} \frac{1}{\lambda\sqrt{\pi}}\exp\left[\frac{A}{\lambda} - \frac{1}{\pi\lambda}\int_0^\infty \frac{2udu}{(1+u^2)^2}\beta\ln(\lambda u) + o\left(\frac{\ln(\lambda)}{\lambda}\right)\right]\end{aligned}$$ (S7)

with $A$ some irrelevant constant

$$\begin{aligned}A(\lambda) &\underset{\lambda\to\infty}{=} \frac{1}{\lambda\sqrt{\pi}}\exp\left[-\beta\frac{\ln(\lambda)}{\pi\lambda} + o\left(\frac{\ln(\lambda)}{\lambda}\right)\right]\\ &\underset{\lambda\to\infty}{=} \frac{1}{\lambda\sqrt{\pi}} - \frac{\beta}{\pi^{\frac{3}{2}}}\frac{\ln(\lambda)}{\lambda^2} + o\left(\frac{\ln(\lambda)}{\lambda^2}\right)\end{aligned}$$

Finally, the inverse Laplace Transform yields the following small $x_0$ behavior :

$$V(x_0) \underset{x_0 \to 0}{=} -\frac{\beta}{\pi^{\frac{3}{2}}} x_0 \ln(x_0) + o\left(x_0 \ln(x_0)\right) \tag{S8}$$

## II. SPLITTING PROBABILITY OF THE LAPLACE JUMP PROCESS

In this section, we show that the splitting probability of the Laplace jump process defined by its jump distribution $f(l) = \frac{1}{2\gamma} e^{-\frac{|l|}{\gamma}}$ satisfies equation (11) of the main text.

### A. Exact determination

The exact calculation can be found in[1]. We here state the result for completeness :

$$\pi_{0,x}(x_0) = \frac{x_0 + \gamma}{x + 2\gamma} \tag{S9}$$

### B. Large $x$ limit and matching with $V(x_0)$

On the one hand, following the main text, the rescaling function reads $A_2(x) = \frac{\gamma\sqrt{\pi}}{x}$ and thus

$$\lim_{x \to \infty} \left[\frac{\pi_{0,x}(x_0)}{A_2(x)}\right] = \frac{1}{\sqrt{\pi}} + \frac{x_0}{\sqrt{\pi}\gamma} \tag{S10}$$

On the other hand, for the Laplace jump process, it is shown in[2] equation (72) that :

$$V(x_0) = \frac{x_0}{\sqrt{\pi}\gamma} \tag{S11}$$

yielding the expected result.

### C. Small $x_0$ behavior

For the Laplace jump process, the Fourier Transform reads $\tilde{f}(k) = \frac{1}{1+\gamma^2 k^2}$. We note that for small $x_0$, $V(x_0)$ is linear in $x_0$ and

$$-\pi^{-\frac{3}{2}} \int_0^\infty \mathrm{d}k \log\left(\frac{\gamma^2 k^2}{1+\gamma^2 k^2}\right) = \frac{1}{\sqrt{\pi}\gamma} \tag{S12}$$

thus confirming the expression of the prefactor.

## III. SPLITTING PROBABILITY OF THE GAMMA JUMP PROCESS

In this section we pursue the same goal as in the previous section for the Gamma Jump Process, defined by its jump distribution $f(l) = \frac{1}{2\gamma^2} |l| e^{-\frac{|l|}{\gamma}}$.

## A. Exact determination

Let us provide the exact expression of the splitting probability for the Gamma jump process. Following the main text, recall that the splitting probability obeys the following integral equation :

$$\pi_{0,\underline{x}}(x_0) = \int_{x-x_0}^{\infty} dx' f(x') + \int_{-x_0}^{x-x_0} dx' \pi_{0,\underline{x}}(x_0 + x') f(x') \tag{S13}$$

and the following differential equation :

$$\frac{d^4}{d^4 x_0} \pi_{0,\underline{x}}(x_0) - \frac{3}{\gamma^2} \frac{d^2}{d^2 x_0} \pi_{0,\underline{x}}(x_0) = 0 \tag{S14}$$

The splitting probability can thus be written as a linear combination of orthogonal solutions :

$$\pi_{0,\underline{x}}(x_0) = A e^{-\frac{\sqrt{3}}{\gamma} x_0} + B e^{+\frac{\sqrt{3}}{\gamma} x_0} + C x_0 + E \tag{S15}$$

with unknown coefficients $A, B, C$ and $E$ depending on $x$. Injecting this form into the integral equation, we obtain 4 independent equations and determine the unknown coefficients and thus the splitting probability:

$$A = \frac{2\gamma e^{\frac{\sqrt{3}x}{\gamma}}}{6\sqrt{3}\gamma - 8\gamma + 2\sqrt{3}x e^{\frac{\sqrt{3}x}{\gamma}} + 3x e^{\frac{\sqrt{3}x}{\gamma}} + 6\sqrt{3}\gamma e^{\frac{\sqrt{3}x}{\gamma}} + 8\gamma e^{\frac{\sqrt{3}x}{\gamma}} + 2\sqrt{3}x - 3x}$$

$$B = \frac{-90\sqrt{3}\gamma - 156\gamma}{108\sqrt{3}\gamma + 186\gamma + 291\sqrt{3}x e^{\frac{\sqrt{3}x}{\gamma}} + 504x e^{\frac{\sqrt{3}x}{\gamma}} + 828\sqrt{3}\gamma e^{\frac{\sqrt{3}x}{\gamma}} + 1434\gamma e^{\frac{\sqrt{3}x}{\gamma}} + 21\sqrt{3}x + 36x}$$

$$\begin{aligned}
C = &\Big[ -189\sqrt{3}\gamma + 327\gamma + 6\sqrt{3}x e^{\frac{2\sqrt{3}x}{\gamma}} + 9x e^{\frac{2\sqrt{3}x}{\gamma}} - 9\sqrt{3}\gamma e^{\frac{\sqrt{3}x}{\gamma}} + 15\gamma e^{\frac{\sqrt{3}x}{\gamma}} + 21\sqrt{3}\gamma e^{\frac{2\sqrt{3}x}{\gamma}} \\
&+ 33\gamma e^{\frac{2\sqrt{3}x}{\gamma}} + 33\sqrt{3}\gamma e^{\frac{3\sqrt{3}x}{\gamma}} + 57\gamma e^{\frac{3\sqrt{3}x}{\gamma}} - 78\sqrt{3}x + 135x \Big] \times \\
&\Big[ -538\sqrt{3}\gamma^2 + 930\gamma^2 + 6\sqrt{3}x^2 e^{\frac{2\sqrt{3}x}{\gamma}} + 9x^2 e^{\frac{2\sqrt{3}x}{\gamma}} - 78\sqrt{3}x^2 + 135x^2 - 54\sqrt{3}\gamma^2 e^{\frac{\sqrt{3}x}{\gamma}} + 102\gamma^2 e^{\frac{\sqrt{3}x}{\gamma}} \\
&+ 66\sqrt{3}\gamma^2 e^{\frac{2\sqrt{3}x}{\gamma}} + 102\gamma^2 e^{\frac{2\sqrt{3}x}{\gamma}} + 94\sqrt{3}\gamma^2 e^{\frac{3\sqrt{3}x}{\gamma}} + 162\gamma^2 e^{\frac{3\sqrt{3}x}{\gamma}} - 21\sqrt{3}\gamma x e^{\frac{\sqrt{3}x}{\gamma}} + 39\gamma x e^{\frac{\sqrt{3}x}{\gamma}} \\
&+ 39\sqrt{3}\gamma x e^{\frac{2\sqrt{3}x}{\gamma}} + 57\gamma x e^{\frac{2\sqrt{3}x}{\gamma}} + 33\sqrt{3}\gamma x e^{\frac{3\sqrt{3}x}{\gamma}} + 57\gamma x e^{\frac{3\sqrt{3}x}{\gamma}} - 411\sqrt{3}\gamma x + 711\gamma x \Big]^{-1}
\end{aligned} \tag{S16}$$

$$\begin{aligned}
E = &\Big[ -66\sqrt{3}\gamma^2 - 66\gamma^2 + 324\sqrt{3}\gamma^2 e^{\frac{\sqrt{3}x}{\gamma}} + 564\gamma^2 e^{\frac{\sqrt{3}x}{\gamma}} + 5574\sqrt{3}\gamma^2 e^{\frac{2\sqrt{3}x}{\gamma}} + 9654\gamma^2 e^{\frac{2\sqrt{3}x}{\gamma}} \\
&+ 114\sqrt{3}\gamma x e^{\frac{\sqrt{3}x}{\gamma}} + 198\gamma x e^{\frac{\sqrt{3}x}{\gamma}} + 1959\sqrt{3}\gamma x e^{\frac{2\sqrt{3}x}{\gamma}} + 3393\gamma x e^{\frac{2\sqrt{3}x}{\gamma}} - 21\sqrt{3}\gamma x - 27\gamma x \Big] \times \\
&\Big[ -132\sqrt{3}\gamma^2 - 132\gamma^2 + 1377\sqrt{3}x^2 e^{\frac{2\sqrt{3}x}{\gamma}} + 2385x^2 e^{\frac{2\sqrt{3}x}{\gamma}} - 9\sqrt{3}x^2 - 9x^2 + 648\sqrt{3}\gamma^2 e^{\frac{\sqrt{3}x}{\gamma}} \\
&+ 1128\gamma^2 e^{\frac{\sqrt{3}x}{\gamma}} + 11148\sqrt{3}\gamma^2 e^{\frac{2\sqrt{3}x}{\gamma}} + 19308\gamma^2 e^{\frac{2\sqrt{3}x}{\gamma}} + 228\sqrt{3}\gamma x e^{\frac{\sqrt{3}x}{\gamma}} + 396\gamma x e^{\frac{\sqrt{3}x}{\gamma}} \\
&+ 7836\sqrt{3}\gamma x e^{\frac{2\sqrt{3}x}{\gamma}} + 13572\gamma x e^{\frac{2\sqrt{3}x}{\gamma}} - 72\sqrt{3}\gamma x - 72\gamma x \Big]^{-1}
\end{aligned}$$

## B. Large $x$ limit and matching with $V(x_0)$

Let us now turn to the large $x$ behavior the splitting probability. For each of the above given coefficients, we obtain the equivalent for $x \to \infty$:

$$A \sim \frac{2\gamma}{2\sqrt{3}x + 3x}$$

$$B \sim \frac{-90\sqrt{3}\gamma - 156\gamma}{291\sqrt{3}xe^{\frac{\sqrt{3}x}{\gamma}} + 504xe^{\frac{\sqrt{3}x}{\gamma}}}$$

$$C \sim \frac{1}{x}$$

$$E \sim \frac{1959\sqrt{3}\gamma + 3393\gamma}{1377\sqrt{3}x + 2385x}$$

(S17)

For this jump process, the rescaling function reads $A_2(x) = \frac{\sqrt{3\pi}\gamma}{x}$ and we thus obtain :

$$\lim_{x \to \infty} \left[ \frac{\pi_{0,x}(x_0)}{A_2(x)} \right] = \frac{1}{\sqrt{\pi}} + \frac{1}{\sqrt{\pi}} \left[ \frac{(\sqrt{3}-1)^2}{3} e^{-\frac{\sqrt{3}}{\gamma}x_0} + \frac{x_0}{\sqrt{3}\gamma} + \frac{2\sqrt{3}-4}{3} \right]$$

(S18)

Finally, for the Gamma jump process, it is shown in[2], equation (77) that :

$$V(x_0) = \frac{1}{\sqrt{\pi}} \left[ \frac{(\sqrt{3}-1)^2}{3} e^{-\frac{\sqrt{3}}{\gamma}x_0} + \frac{x_0}{\sqrt{3}\gamma} + \frac{2\sqrt{3}-4}{3} \right]$$

(S19)

yielding the agreement with equation (11) of the main text.

### C. Small $x_0$ behavior

For the Gamma jump process, the Fourier Transform reads $\tilde{f}(k) = \frac{1-k^2\gamma^2}{(1+\gamma^2 k^2)^2}$. We note that for small $x_0$, $V(x_0)$ is linear in $x_0$ and

$$V(x_0) \underset{x_0 \to 0}{=} \frac{2-\sqrt{3}}{\sqrt{\pi}} \frac{x_0}{\gamma} + o(x_0)$$

(S20)

One can show as well that

$$-\pi^{-\frac{3}{2}} \int_0^\infty dk \log\left(1 - \frac{1-k^2\gamma^2}{(1+\gamma^2 k^2)^2}\right) = \frac{2-\sqrt{3}}{\sqrt{\pi}} \frac{x_0}{\gamma}$$

(S21)

thus confirming the expression of the prefactor in the small $x_0$ limit and agreeing with equation (13) of the main text.

### IV. FOURIER TRANSFORMED JUMP DISTRIBUTIONS FOR VARIOUS GAMMA JUMP PROCESSES

We recall that the jump distribution reads, for $n > -1$:

$$f(l) = \frac{1}{2\gamma^{n+1}\Gamma(n+1)} |l|^n e^{-\frac{|l|}{\gamma}}.$$

(S22)

Let us give the Fourier transforms for the various values of $n$ used in the main text:

- $n = 0$

$$\tilde{f}_0(k) = \frac{1}{1+\gamma^2 k^2} \tag{S23}$$

- $n = 2$

$$\tilde{f}_2(k) = \frac{1 - 3k^2\gamma^2}{(1+\gamma^2 k^2)^3} \tag{S24}$$

- $n = -1/2$

$$\tilde{f}_{-\frac{1}{2}}(k) = \sqrt{\frac{1+\sqrt{1+k^2\gamma^2}}{2+2\gamma^2 k^2}} \tag{S25}$$

From these expressions, one easily obtains the large and small $k$ behavior of the Fourier Transform and thus the transition probabilities and small $x_0$ asymptotic limit of the splitting probabilities for these Gamma jump processes, as claimed in the main text.

## V. EFFECTIVE $1d$ JUMP DISTRIBUTION IN $3d$ GEOMETRIES

In this section, we provide the derivation of the effective $1d$ jump distribution in the $3d$ slab geometry. Let us first remind that for the jump processes considered, the projection on one coordinate of the jump distribution reads

$$f(l) = \frac{1}{2}\int_{|l|}^\infty \frac{p(r)}{r}\mathrm{d}r \tag{S26}$$

with $p(r)$ the distribution of the module, the angle being distributed uniformly on the unit sphere.

### A. Exponential distribution

We first consider the case where the module of the jump is distributed according to $p(r) = \frac{1}{\gamma}e^{-\frac{r}{\gamma}}$. One then has:

$$\begin{aligned} f(l) &= \frac{1}{2}\int_{|l|}^\infty \frac{p(r)}{r}\mathrm{d}r \\ &= \frac{1}{2\gamma}\int_{|l|}^\infty \frac{1}{r}e^{-\frac{r}{\gamma}}\mathrm{d}r \\ &= \frac{1}{2\gamma}\int_{\frac{|l|}{\gamma}}^\infty \frac{1}{r}e^{-r}\mathrm{d}r \\ f(l) &= \frac{1}{2\gamma}\Gamma\left(0, \frac{|l|}{\gamma}\right) \end{aligned} \tag{S27}$$

and

$$\tilde{f}(k) = \frac{\arctan(k\gamma)}{k\gamma} \tag{S28}$$

The large and small $k$ expansions read:

$$\begin{aligned} \tilde{f}(k) &\underset{k\to 0}{=} 1 - \frac{\gamma^2 k^2}{3} + o(k^2) \\ \tilde{f}(k) &\underset{k\to\infty}{=} \frac{\pi}{2\gamma k} + o(k^{-1}) \end{aligned} \tag{S29}$$

yielding $A_2 = \sqrt{\frac{\pi}{3}}\frac{\gamma}{x}$ and thus the following small $x_0$ behavior :

$$\pi_{0,\underline{x}}(x_0) \underset{x\to\infty}{\sim} \frac{1}{\sqrt{3}x}\left[\gamma - \frac{x_0\ln(x_0)}{2} + o\left(x_0\ln(x_0)\right)\right], \tag{S30}$$

as claimed in the main text.

### B.  $\alpha$-stable jump process

We now turn to an $\alpha$-stable jump process with module distributed as :

$$\begin{aligned} p(r) &= p_1(r) + p_1(-r) \\ \text{with} \quad \tilde{p_1}(k) &= e^{-|a_\mu k|^\mu} \end{aligned} \tag{S31}$$

Let us directly compute the Fourier Transform of the effective $1d$ distribution:

$$\begin{aligned} \tilde{f}(k) &= \frac{1}{2}\int_{-\infty}^\infty e^{ikl}\left[\int_0^\infty \frac{1}{r}\theta\left(1-\left|\frac{r}{l}\right|\right)(p_1(r)+p_1(-r))\mathrm{d}r\right]\mathrm{d}l \\ &= \int_0^\infty \frac{\sin(kr)}{kr}(p_1(r)+p_1(-r))\,\mathrm{d}r \\ &= \frac{1}{k}\int_{-\infty}^\infty \frac{\sin(kr)}{r}p_1(r)\mathrm{d}r \end{aligned} \tag{S32}$$

Let us now go into the complex formalism, and we will take the imaginary part at the end:

$$\begin{aligned} \tilde{f}(k) &= \frac{1}{k}\int_{-\infty}^\infty \frac{e^{ikr}}{r}p_1(r)\mathrm{d}l \\ &= \frac{1}{k}\int_{-\infty}^\infty \frac{e^{ikr}}{r}\left[\frac{1}{2\pi}\int_{-\infty}^\infty e^{-ik'r-|a_\mu k'|^\mu}\mathrm{d}k'\right]\mathrm{d}r \\ &= \frac{1}{k}\int_{-\infty}^\infty \left[\frac{1}{2\pi}\int_{-\infty}^\infty \frac{e^{ir(k-k')}}{r}\mathrm{d}r\right]e^{-|a_\mu k'|^\mu}\mathrm{d}k' \\ &= \frac{i}{k}\int_{-\infty}^\infty \frac{1}{2}\mathrm{sgn}(k-k')e^{-|a_\mu k'|^\mu}\mathrm{d}k' \end{aligned} \tag{S33}$$

Finally :

$$\tilde{f}(k) = \frac{\Gamma\left(\frac{1}{\mu}\right) - \Gamma\left(\frac{1}{\mu},(a_\mu k)^\mu\right)}{a_\mu \mu k}, \tag{S34}$$

and the large and small $k$ expansions read:

$$\begin{aligned} \tilde{f}(k) &\underset{k\to 0}{\sim} 1 - \frac{(a_\mu k)^\mu}{1+\mu} + o(k^\mu), \\ \tilde{f}(k) &\underset{k\to\infty}{\sim} \frac{1}{k}\frac{\Gamma\left(\mu^{-1}\right)}{a_\mu \mu} + o(k^{-1}), \end{aligned} \tag{S35}$$

yielding result (23) of the main text.

---

[1] Van N.G. Kampen. *Stochastic Processes in Physics and Chemistry*. Elsevier, 2007.
[2] Satya N. Majumdar, Philippe Mounaix, and Grégory Schehr. Survival probability of random walks and Lévy flights on a semi-infinite line. *Journal of Physics A: Mathematical and Theoretical*, 50(46):465002, nov 2017.



\end{document}